\documentclass[a4paper,fleqn,usenatbib]{mnras}
\usepackage{newtxtext,newtxmath}
\usepackage[T1]{fontenc}
\usepackage{ae,aecompl}

\usepackage{graphicx}	
\usepackage{amsmath}	
\usepackage{amssymb}	
\usepackage{verbatim,graphicx,epsfig,dcolumn,xcolor}

\newcolumntype{.}{D{.}{.}{4}}
\newcolumntype{,}{D{.}{.}{2}}
\newcolumntype{;}{D{.}{.}{1}}

\newcommand{\appropto}{\mathrel{\vcenter{
  \offinterlineskip\halign{\hfil$##$\cr
    \propto\cr\noalign{\kern2pt}\sim\cr\noalign{\kern-2pt}}}}}

\title[AGB star winds in the LMC]{The onset of the AGB wind tied to a transition between sequences in the period--luminosity diagram}
\author[I. McDonald \& M. Trabucchi]{
I.~McDonald$^{1}$\thanks{E-mail: iain.mcdonald-2@manchester.ac.uk},
M.~Trabucchi$^{2}$
\\
$^{1}$Jodrell Bank Centre for Astrophysics, Alan Turing Building, Manchester, M13 9PL, UK\\
$^{2}$Dipartimento di Fisica e Astronomia Galileo Galilei Università di Padova, Vicolo dell'Osservatorio 3, I-35122 Padova, Italy
}

\date{Accepted XXX. Received YYY; in original form ZZZ}

\pubyear{2018}

\begin{document}
\label{firstpage}
\pagerange{\pageref{firstpage}--\pageref{lastpage}}
\maketitle

\begin{abstract}
We link the onset of pulsation-enhanced, dust-driven winds from asymptotic giant branch (AGB) stars in the Magellanic Clouds to the star's transition between period--luminosity sequences (from $B$ to $C^\prime$). This transition occurs at $\sim$60 days for {solar-mass stars, which represent the bulk of the AGB population: this is} the same period at which copious dust production starts in solar-neighbourhood AGB stars. It is contemporaneous with the onset of long-secondary period (LSP) variability on sequence $D$. The combined amplitude of the first-overtone ($B+C^\prime$) and fundamental ($C$) modes and (perhaps) long-secondary period ($D$; LSP) variability appears to drive a sudden increase in mass-loss rate to a stable plateau, previously identified to be a few $\times$ 10$^{-7}$ M$_\odot$ yr$^{-1}$. We cite this as evidence that pulsations are necessary to initiate mass loss from AGB stars and that these pulsations are significant in controlling stars' mass-loss rates. We also show evidence that LSPs may evolve from long to short periods as the star evolves, counter to the other period--luminosity sequences.
\end{abstract}




\section{Introduction}
\label{IntroSect}

Mass loss from asymptotic giant branch (AGB) stars controls the last stages of a star's life \citep[e.g.][]{KL14}. Here, pulsations levitate material from the stellar surface, where it can condense into dust. Radiation pressure on this dust forces it from the star, and collisional coupling to the surrounding gas drives a wind from the star \citep[e.g.][]{Willson00,HO18}. The conditions required to initiate this process, and the mass-loss rates it generates, are poorly known.

As well as being significant for modelling stellar evolution and galactic chemical enrichment, this becomes significant in metal-poor stars: if pulsation starts the wind and sets the mass-loss rate, the mass-loss rate should be largely independent of metallicity; but if radiation pressure does, it should be highly metallicity dependant. Definitive observations do not currently allow this distinction to be made: in our limited observational scope, metallicity seems to make no difference to mass-loss rate \citep{vLBM08} and dust-production seems prevalent in metal-poor environments \citep[e.g.][]{BMG+17}. However, a decrease in wind velocity may exist at lower metallicities \citep{GVM+16,MSS+16,GvLZ+17}. \citet{MBG+19} interpret this as being due to the levitating effects of pulsations setting the mass-loss rate, and radiation pressure on dust setting the wind momentum. However, this hypothesis does not predict a mass-loss rate for individual stars, nor determine which stars do and do not produce a strong, dust-driven wind.

The link between pulsation and mass loss has a long history: stars with significant infrared excess (indicating high mass-loss rates) are typically highly variable (strongly pulsating), long-period stars \citep[e.g.][]{Cannon67,Habing96}. Most studies \citep[e.g.][]{VW93} have concentrated on long-period ($P \gtrsim 300$ days), large-amplitude ($\delta V > 2.5$ mag) Mira variables. However, at least in Galactic stars, strong mass loss ($\dot{M} \gtrsim 10^{-8}$ M$_\odot$ yr$^{-1}$) first occurs around a pulsation period of $P \sim 60$ days, as traced by both infrared excess (indicating dust) and CO line-profile data \citep{ABC+01,GvL07,GSB+09,MZ16,MDBZL18}. To the limits of modern detection, mass loss identified by these tracers is always linked to optical variability, hence pulsation is a necessary condition for stellar dust production and associated gas mass loss \citep{Groenewegen14,MZB12,MZS+16,MZW17}.

Properties of variable stars are often traced through the period--luminosity ($P-L$) diagram \citep[e.g.][]{ITM+04,Wood15}, in which stars form distinct sequences depending on the pulsation mode responsible for their variability. Pulsating red giants are often multi-periodic, and normally only the primary period (the one with the largest amplitude) in each star is used in the $P-L$ diagram. As stars evolve, specific overtone modes gradually become stable, and the primary mode shifts towards lower radial orders \citep[e.g.][]{LW03,TWM+19}. Sequence $C$, where Mira variables are generally located, is due to pulsation in the fundamental mode \citep[e.g.][]{Wood15}. \citet{TWM+17} have recently shown that both sequences $B$ and $C^\prime$ are due to pulsation in the first overtone (1O) mode, while sequences $A$ and $A^\prime$ are due to the second (2O) and third overtone (3O) modes, respectively. Sequence $D$ is associated with long-secondary periods (LSPs), a mode of variability with unknown origin \citep[e.g.][]{NWCS09}, but whose onset is linked by \citet{TWM+17} to the gap between sequences $B$ and $C^\prime$. Stars initially rise on the $P-L$ diagram while traversing these sequences from left to right but, as increasing mass loss decreases the density of the stellar envelope, their tracks tend to flatten towards the end of the AGB evolution \citep[e.g.][their figure 20]{VW93}.

$P \sim 60$ days approximately corresponds to the lower end of pulsation sequence $C^\prime$, hence the onset of mass loss at this stage may be linked to both the start of this sequence and/or the onset of LSPs. The link between LSPs and mass loss via dust-driven winds was first raised by \citet{WN09}, who suggested that mass may be lost via a clumpy or disc-like structure. However, while most stars appears to start strong mass loss at this period, not all do: a factor \citet{MDBZL18} suggest is linked to stellar mass.

The Large Magellanic Cloud (LMC) offers a well-studied environment with a diverse range of stellar populations at approximately half of solar metallicity. Mass-loss rates for AGB stars have been estimated \citep{RSSM12}, based on their infrared colours \citep{MGI+06,SSM11}, and pulsation periods of these stars are also known \citep[e.g.][]{FHCK05,DKB+06,SUS+09}. Unlike Galactic stars, they reside at a single, well-known distance, making construction of a $P-L$ diagram trivial (Figure \ref{PLFig}). In this Letter, we connect the $P-L$ diagram of the LMC to the mass-loss rates of its stars, in order to explore the evolutionary connection between mass loss and pulsation properties.


\section{Results}

\begin{figure*}
\centerline{\includegraphics[width=0.35\textwidth,angle=-90]{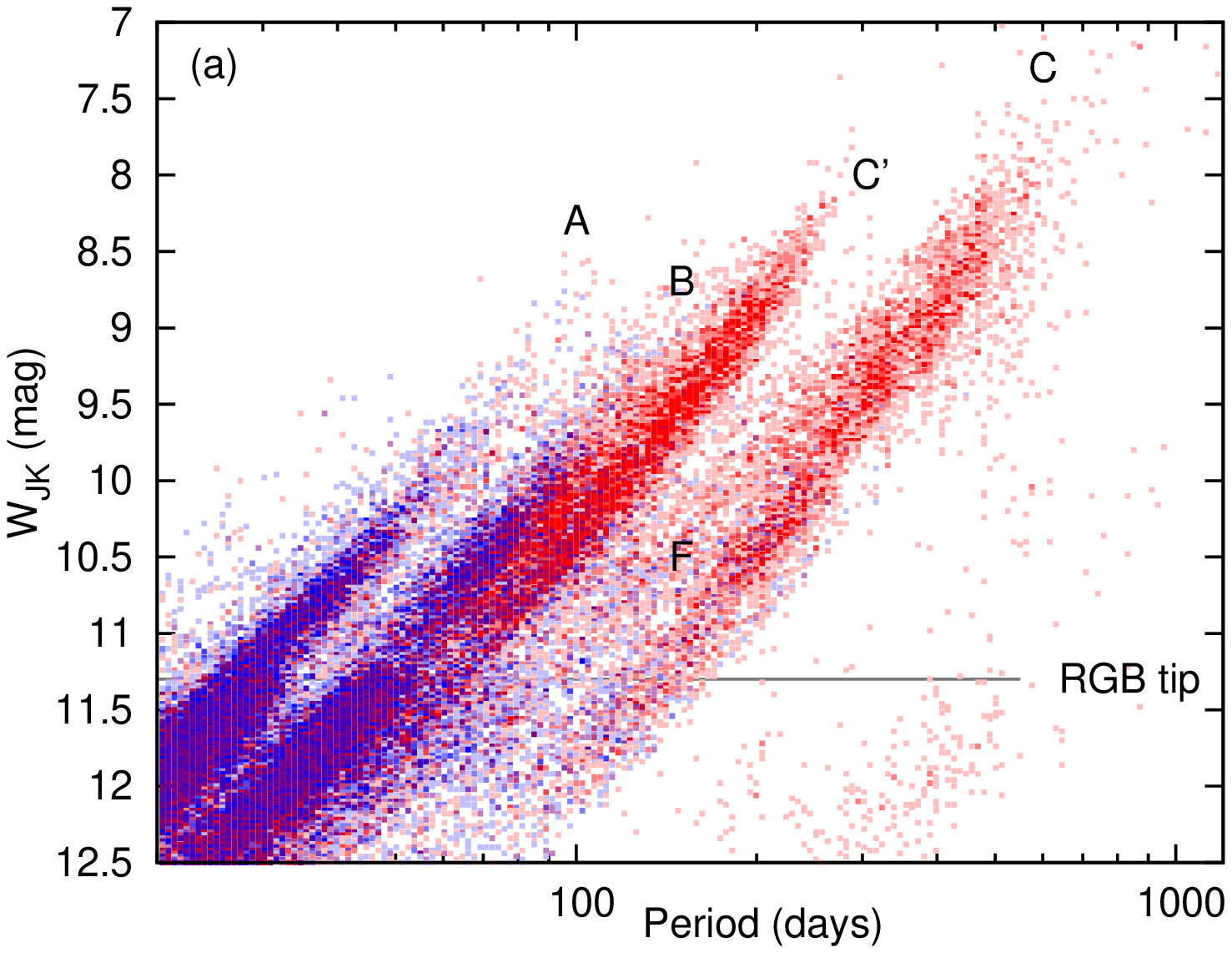}
            \includegraphics[width=0.35\textwidth,angle=-90]{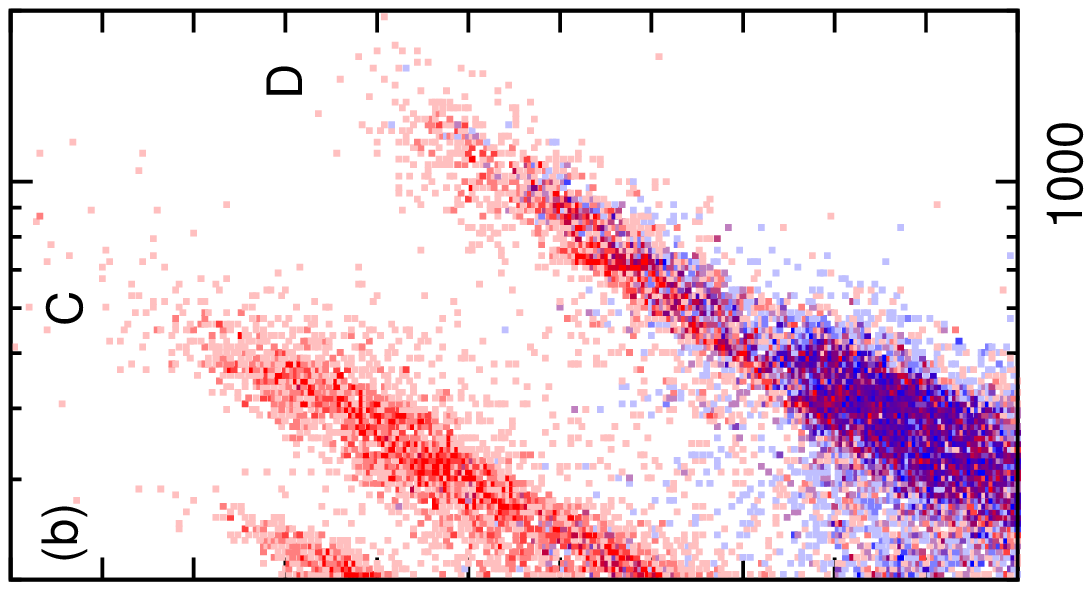}
            \includegraphics[width=0.35\textwidth,angle=-90]{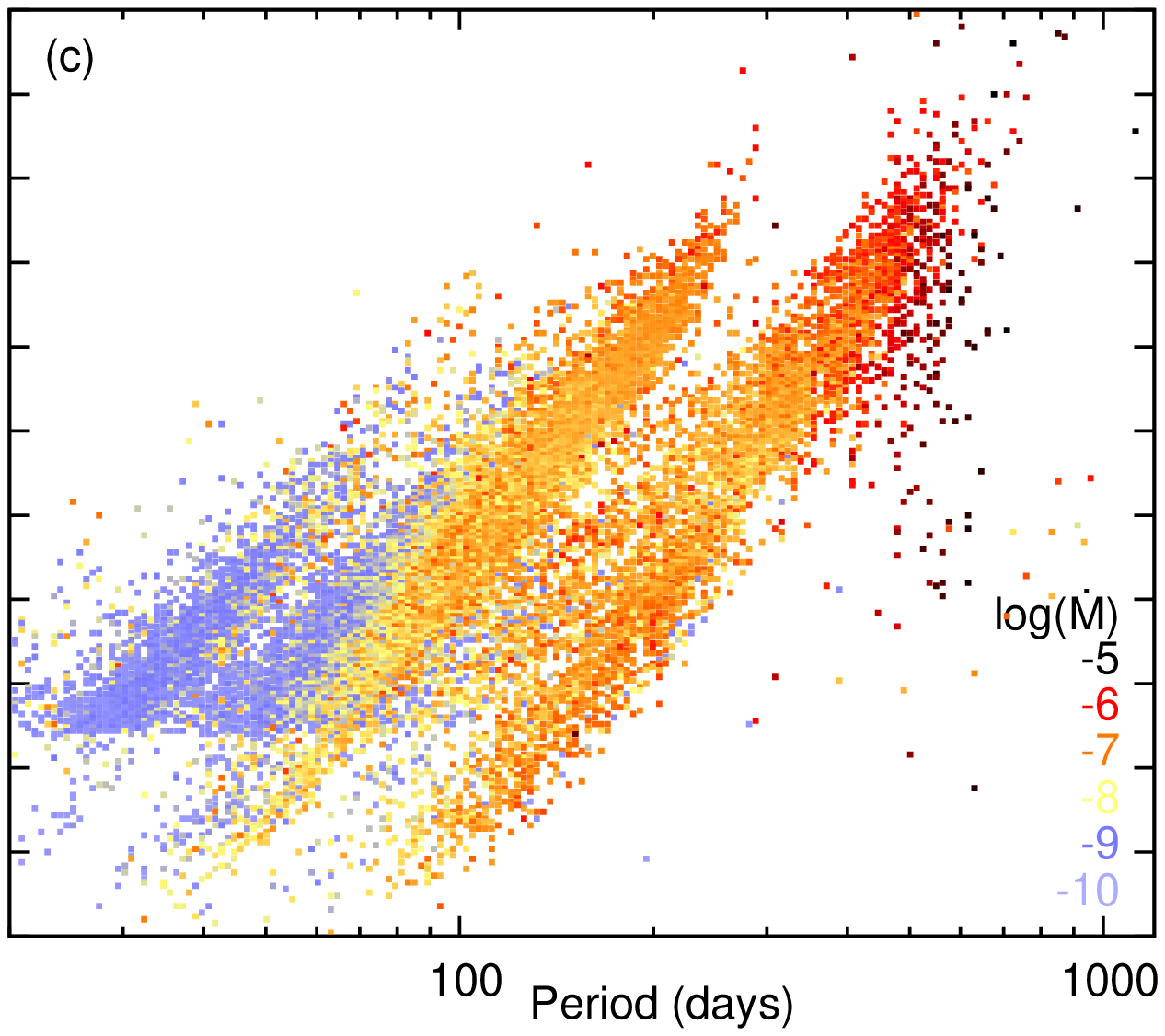}}
\caption{Binned period--luminosity diagrams. Sequences are labelled according to the nomenclature used in \citet{Wood15}. Panel (a): fraction of stars with infrared excess ($K-[11.3]>0.3$ mag), from none (blue) to all (red), with sequence $D$ variables remapped; (b) sequence $D$ variables in situ; (c) as (a), but colour-coded by the average log($\dot{M}$) from \citet{RSSM12}, assuming a gas:dust ratio of 400.}
\label{PLFig}
\end{figure*}

\begin{figure*}
\centerline{\includegraphics[width=0.35\textwidth,angle=-90]{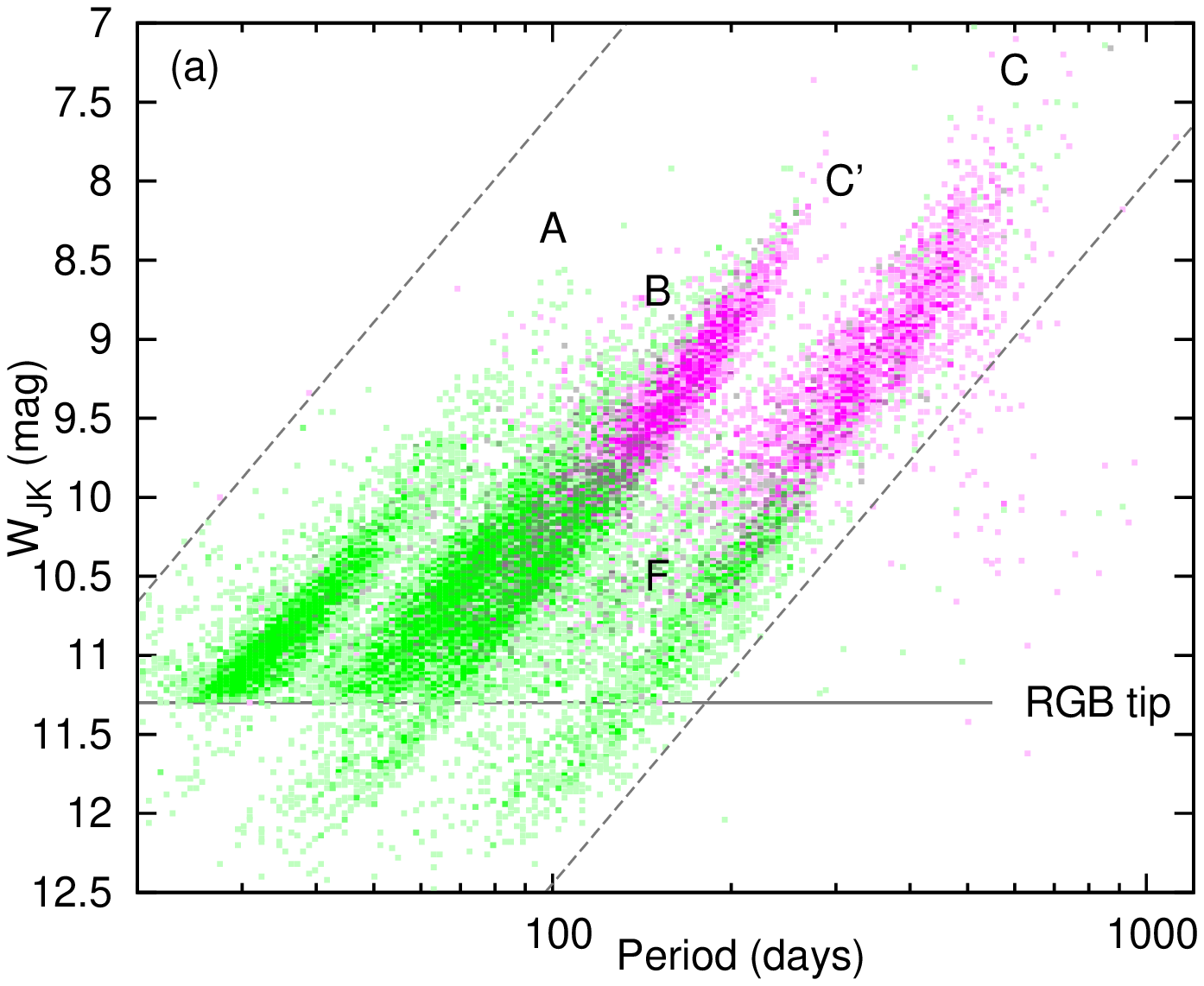}
            \includegraphics[width=0.35\textwidth,angle=-90]{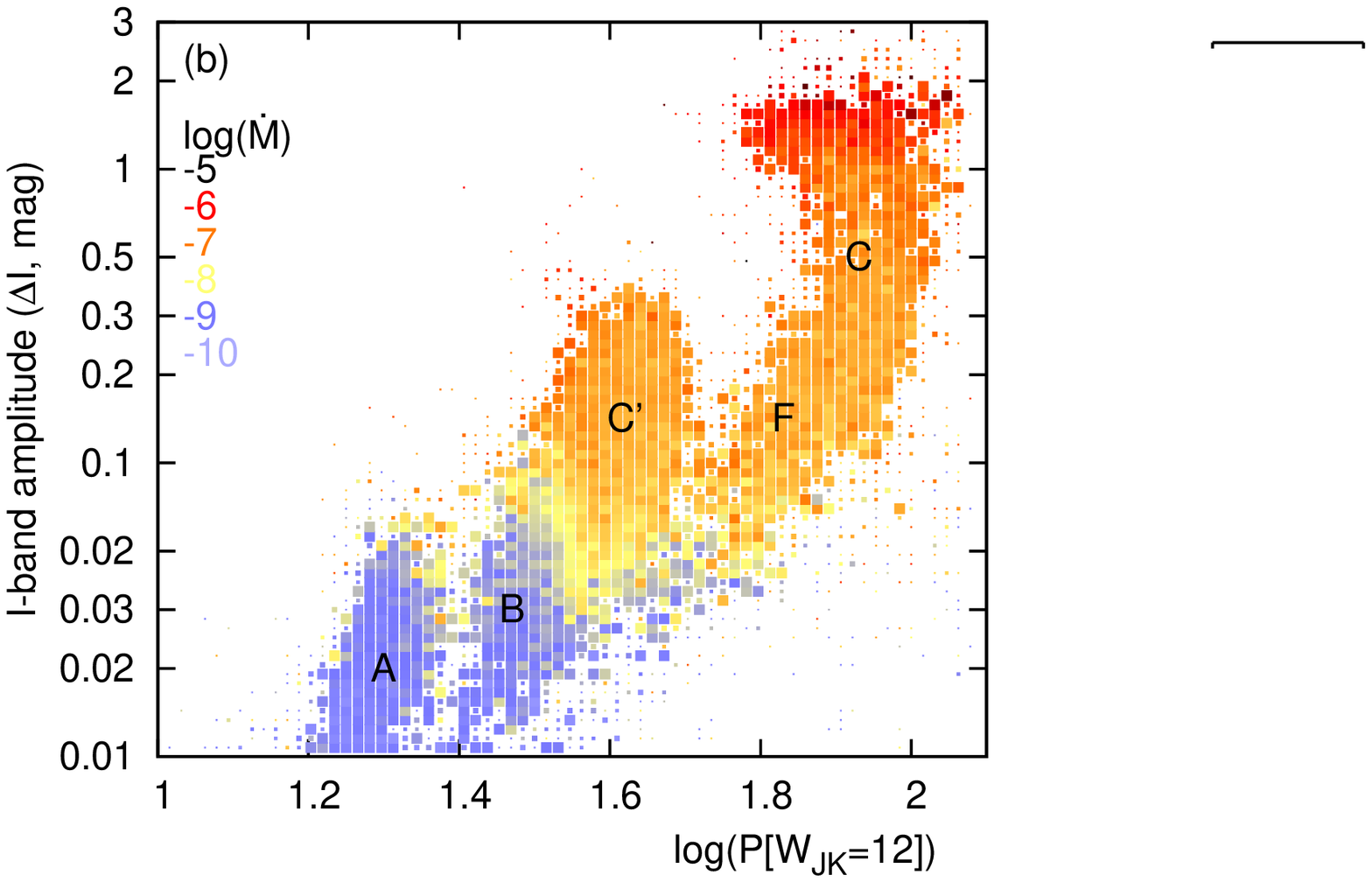}
            \includegraphics[width=0.35\textwidth,angle=-90]{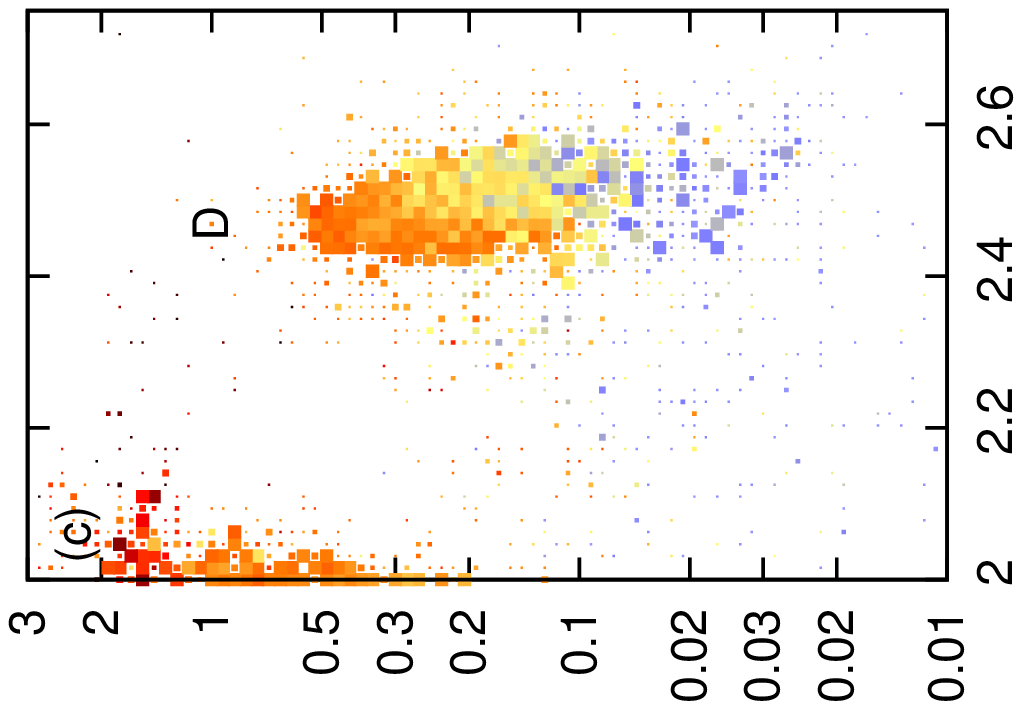}}
\caption{Panel (a): binned period--luminosity diagram, showing the fraction of stars identified as having carbon-based or high-opacity dust (C-AGB or X-AGB classifications in \citet{RSSM12}), from none (green) to all (purple), with sequence $D$ variables remapped. {Dashed, diagonal lines show the projection used to construct the horizontal axis of panels (b) and (c), as described in the text.} Panels (b) and (c) show the projected pulsation mode versus pulsation amplitude, coloured by mass-loss rate. Sequence $D$ variables are remapped in panel (b) and in situ in panel (c). The second-overtone ($A$), first-overtone ($B$, $C\prime$), and fundamental-mode ($C$ and possibly $F$) pulsation sequences are marked.}
\label{PAFig}
\end{figure*}

Pulsation periods and $I$-band amplitudes ($\Delta I$) were retrieved from the Optical Gravitational Lensing Experiment (OGLE; \citep{SUK+04}) Catalogue of Long-Period Variables in the LMC \citep{SUS+09}. This database was cross-correlated with \citet{RSSM12} to obtain broadband photometry and mass-loss rates for each star, and with the updated \emph{Wide-field Infrared Survey Explorer} (\emph{AllWise}; \citep{CWC+13}) catalogue, to add further infrared photometry. We multiply the mass-loss rates from \citet{RSSM12} by a gas:dust ratio of 400, to approximate the total mass-loss rate. Mass-loss rates are not available for stars below the RGB tip ($K_{\rm s} = 12$ mag), unless they have notable excess infrared flux in \emph{Spitzer}. The \emph{Spitzer} 24 $\mu$m and \emph{WISE} 22 $\mu$m photometry is insufficiently sensitive to measure most stars, hence we use \emph{WISE} 11.3 $\mu$m magnitudes to indicate infrared excess, using the criterion $K_{\rm s}-[11.3] > 0.3$ mag to define stars as dust-producing\footnote{This approximates the same set of stars for which $K_{\rm s}-[22] > 0.55$ mag, the criterion used by \citet{MDBZL18}.}. We also calculate the reddening-free index, $W_{JK} = K_{\rm s} - 0.686(J-K_{\rm s})$ \citep[cf.][]{Madore82,SUK+05}.

\citet{TWM+17} found that primary periods on both sequences $B$ and $C^\prime$ are due to pulsation in the 1O mode. Stars transitioning from sequence $B$ to sequence $C^\prime$ develop a LSP with a larger photometric amplitude than that of the 1O mode. The LSP becomes classified as primary period, resulting in an apparent gap between sequences $B$ and $C^\prime$. In the following, if the primary observed mode lies on sequence $D$, we replace it by the secondary observed mode. 
We do not correct ``extreme'' AGB stars with $J-K_{\rm s} > 3$ mag, as these are likely to intrinsically be on sequence $C$, but be self-extincted by their own dust in $K_{\rm s}$.

Figure \ref{PLFig}(a) shows the $P-L$ diagram, with LSPs remapped in this manner, colour-coded by the fraction of stars in each bin meeting our dust-producing criterion. A striking difference is apparent between the stars on sequences $A$ and $B$, where very few stars exhibit dust production, and sequences $C^\prime$ and $C$, where almost all stars exhibit strong dust production. This is also apparent in the total estimated mass-loss rates (Figure \ref{PLFig}(c)), which are typically $\sim$10$^{-9}$ M$_\odot$ yr$^{-1}$ on sequences $A$ and $B$, and $\sim$10$^{-7}$ M$_\odot$ yr$^{-1}$ on sequences $C^\prime$ and $C$. Stars on sequence $F$, which is of unknown origin but may be a fundamental mode pulsation \citep{SW13,Wood15}, also show dust.

Figure \ref{PLFig}(b) shows the same diagram, but including only primary or secondary periods which fall into the LSP regime. Below the RGB tip, the fraction of dusty stars is almost zero, while above $W_{JK} \approx 9.5$ mag, all stars are dust-producing. Between these magnitudes, stars on the long-period side of the LSP sequence produce little dust, while those on the short-period side do. This is opposite to the typical direction (i.e., longer-period stars typically produce more dust).

The picture is complicated by the formation of carbon stars, due to convective dredge-up of nuclear-processed material with a high carbon abundance during thermal pulses \citep{Herwig05}. This changes the dust chemistry from oxygen- to carbon-rich, resulting in substantially different dust spectra and can affect the amount of infrared excess that a given mass-loss rate will produce \citep[e.g][]{JWK+17}. \citet{RSSM12} fit most sources with carbon-rich dust above $W_{JK} \approx 10$ (Figure \ref{PAFig}(a)), but are not able to reliably constrain carbon-richness in sources not producing substantial amounts of dust. The primary differences between sequences $B$ and $C^\prime$ are fainter than this transition, and remain when identified carbon stars are deselected. This transition to dust production therefore does not appear strongly linked to the phenomenon of convective dredge-up.

Figures \ref{PAFig}(b) and (c) show a variant of the period--amplitude diagram constructed with OGLE data \citep[e.g.][their figure 3]{SUS+09}. The period on the horizontal axis is replaced with the parameter $\log(P[W_{JK}=12])=\log P+(W_{JK}-12)/4.444$, introduced by \citep{Wood15}. This quantity is obtained by projecting $\log P$ along a line roughly parallel to the P-L sequences until it meets the level $W_{JK}=12$. In other words, the sequences are ``skewed'' and turned into vertical distributions, so that two stars with primary modes on the same sequence will have similar values of $\log(P[W_{JK}=12])$, regardless of their brightness. This diagram allows to clearly distinguish between periods associated with different pulsation modes (cf.\ \citealt{TWM+17}, their figure 7). Note that LSPs have been remapped in Figure \ref{PAFig}(b) and retained in Figure \ref{PAFig}(c). A clear correlation exists whereby stars that have large $\Delta I$ tend to have a significantly higher mass-loss rate. These objects pulsate primarily in the fundamental mode or 1O mode. However, once $\Delta I \gtrsim 0.1$ mag, the mass-loss rate stops increasing, reaching a stable value of $\sim10^{-7}\,{\rm M}_{\odot}\,{\rm yr}^{-1}$, and resumes increasing only when $\Delta I \gtrsim 1$ mag (see also Figure \ref{DiscFig}(c)).


\section{Discussion}

\begin{figure*}
\centerline{\includegraphics[width=0.33\textwidth,angle=-90]{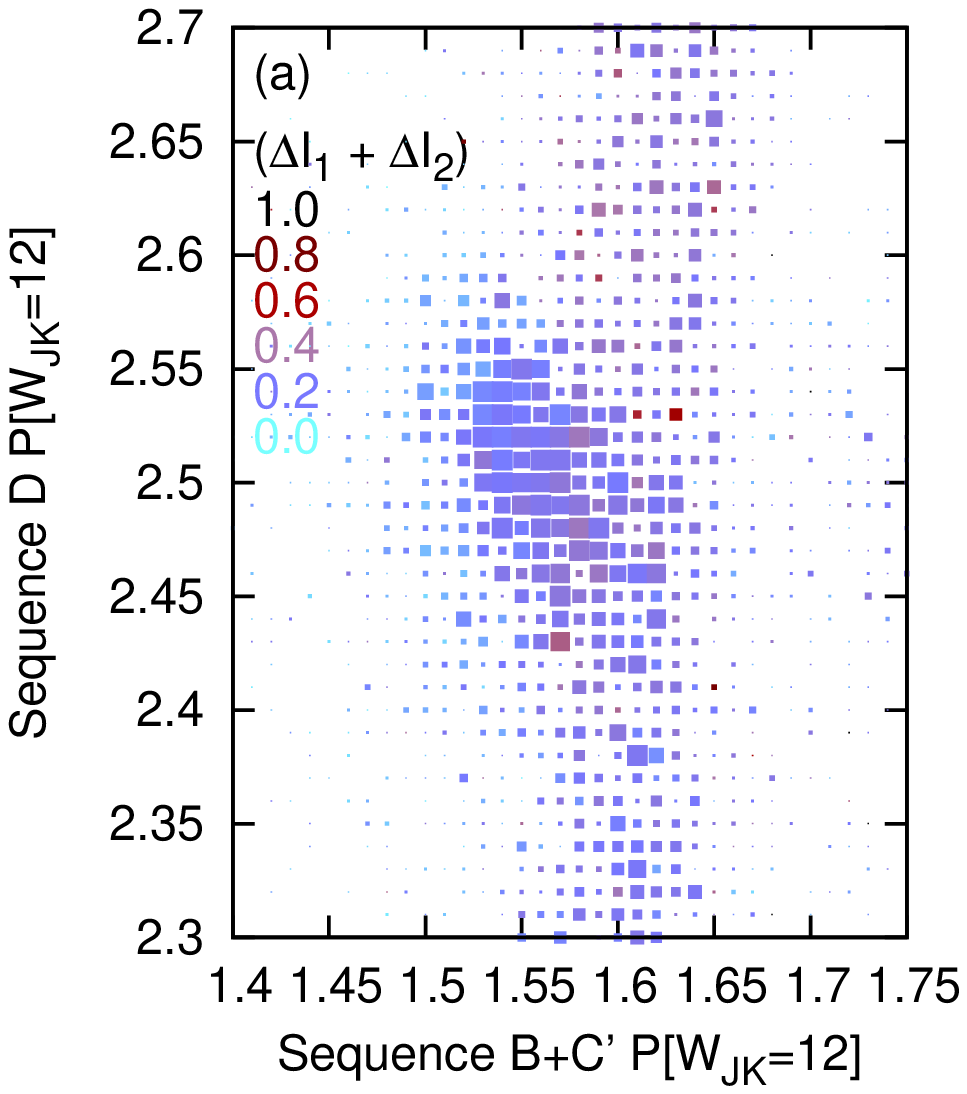}
            \includegraphics[width=0.33\textwidth,angle=-90]{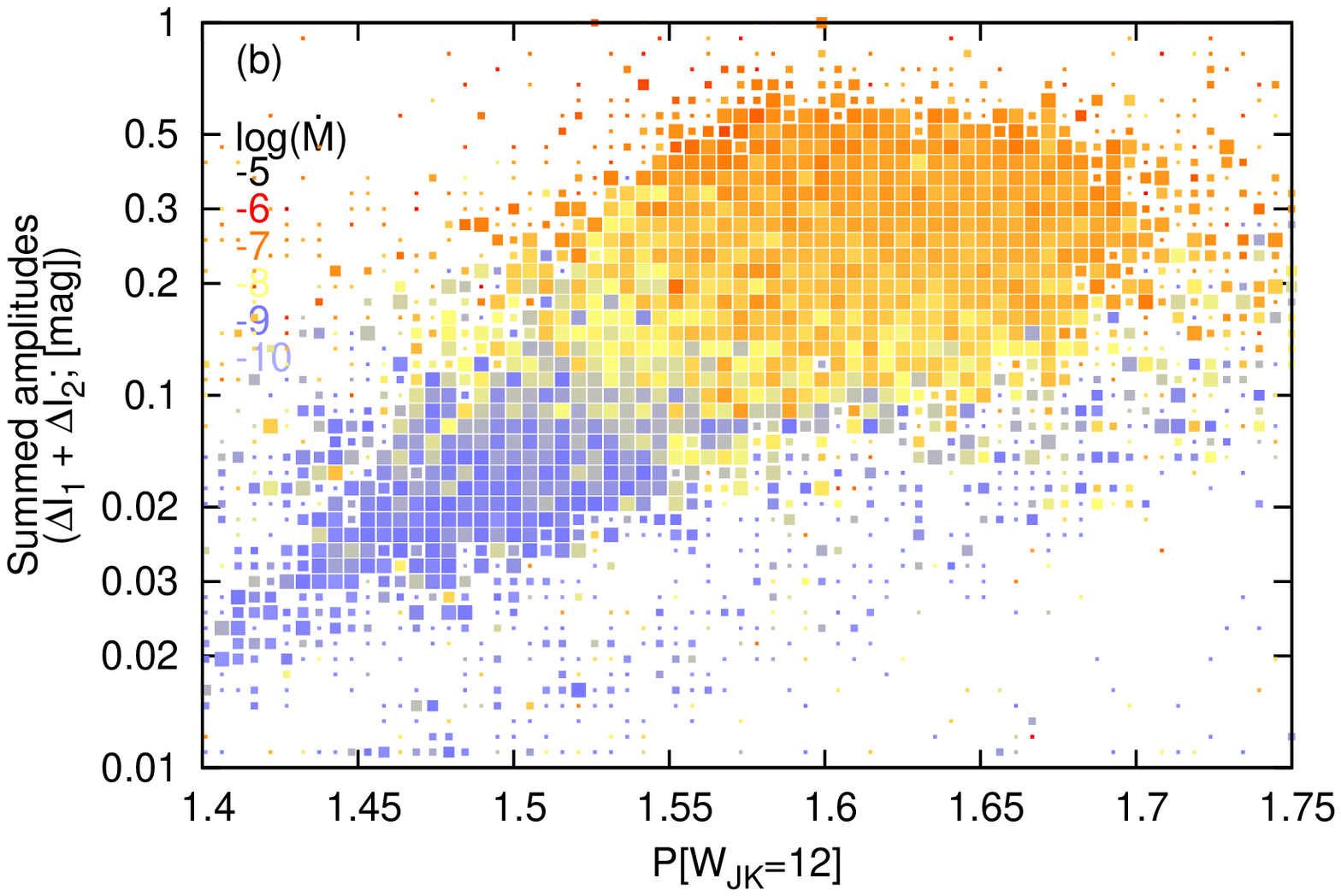}
            \includegraphics[width=0.33\textwidth,angle=-90]{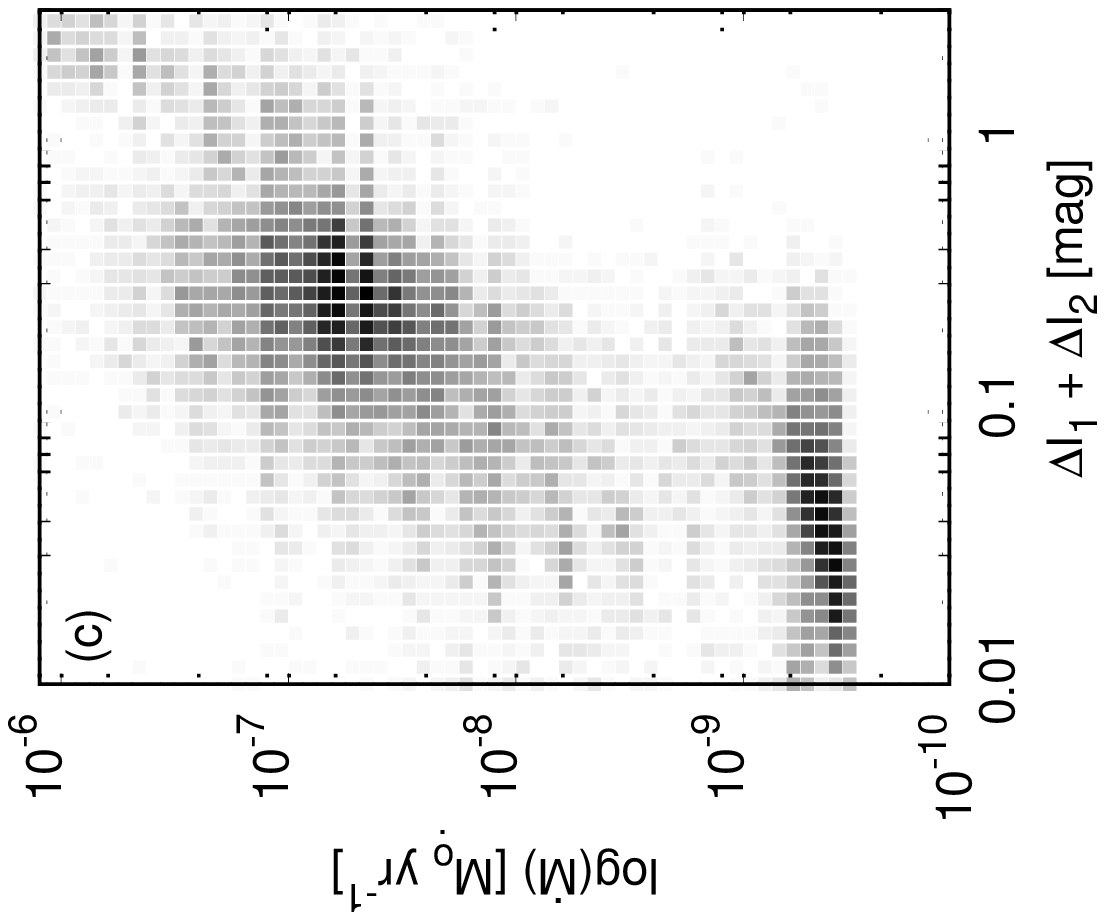}}
\caption{Panel (a): unbinned, projected periods ($P[W_{JK}=12]$) in the 1O mode, versus LSP periods, colour-coded by summed primary- and second-pulsation-period amplitudes. Panel (b): projected period on sequences $B$ and $C^\prime$, versus the summed amplitudes. Panel (c): density plot showing the relation between the summed amplitudes and mass-loss rate.}
\label{DiscFig}
\end{figure*}

A clear link exists between four evolutionary changes: the transition from sequences $B$ to $C^\prime$, an increase in the photometric amplitude of the 1O pulsation mode (a larger $\Delta I$), the development of LSPs, and the onset of substantial mass loss ($\dot{M} \gtrsim 10^{-8}$ M$_\odot$ yr$^{-1}$). The conditions for achieving $\dot{M} \gtrsim 10^{-8}$ M$_\odot$ yr$^{-1}$ appear twofold: firstly, the star has to traverse from sequence $B$ to $C^\prime$ ($\log(P[W_{JK}=12]) \gtrsim 1.55$ dex), and it has to attain an amplitude of $\Delta I \gtrsim 0.05$ mag in either the 1O or fundamental-mode pulsation. If either of these criteria is not met, strong mass loss does not occur.

The transition from $B$ to $C^\prime$ occurs at different periods, dependent predominantly on stellar mass. Sequence $C^\prime$ starts to become populated at $P \approx 40$ days. Sequence $B$ effectively terminates at $P \approx 100$ days, hence most stars should have transitioned to sequence $C^\prime$ by this point. Strong dust production is exhibited on sequence $C^\prime$ at all periods across this range. In order to occupy the shortest-period end of sequence $C^\prime$ ($\approx$40 days), stars must be of very low mass. Hence, also likely very old and probably metal-poor\footnote{We remind the reader that metal-poor stars evolve faster, hence AGB stars can extend to lower masses.}. Sequence $C^\prime$ stars at this period and luminosity are rare, but an analogy can be drawn with 47 Tuc V13 ($P = 40$ days; $W_{JK}$ = 12.14 mag the distance of the LMC; $K-[11.3]$ = 0.907 mag; \citep{MBvLZ11}). The rarity of such stars, and lack of an increase in sequence $C^\prime$ stars below the RGB tip, likely indicate that even the lowest-mass RGB stars ($\sim$0.65 M$_\odot$; cf.\ \citealt{MZ15b}) are too massive to exhibit strong pulsations on this sequence, and that the mass-losing stars on all sequences are restricted exclusively to AGB stars.

Sequence $C^\prime$ becomes well-populated by $P \approx 60$ days. This can be compared with Galactic observations, where significant dust production and mass loss is generally only found in stars of $P \gtrsim 60$ days. Significant mass loss becomes essentially ubiquitous at $P \approx 100$ days \citep{MZ16,MDBZL18}. A picture then emerges tying these key periods to changes in the population density of sequence $C^\prime$, and that it is presence on this sequence, rather than pulsation period \emph{per se}, that dictates whether a star undergoes strong mass loss. An interesting feature of this is that the periods ($\sim$60 and 100 days) at which strong mass loss begins are very similar between the LMC and the Galaxy, despite the differences in metallicity and history between the two systems.

The causative mechanism(s) behind the sudden onset of mass loss, however, remain more elusive. Not all sequence $B+C^\prime$ stars with LSPs exhibit dust production, even above the RGB tip (Figure \ref{PLFig}(b)), hence the growth of the LSP at the start of this transition may precede the onset of strong mass loss. The increasing LSP amplitude is closely correlated with the $B+C^\prime$ amplitude. There appears to be no particular pulsation amplitude on sequences $B+C^\prime$ or LSP sequence $D$ where strong mass loss is triggered. Instead, strong mass loss more clearly begins at a given period than at a given amplitude ($\log{P_[W_{JK}=12]}=1.54$ dex; Figure \ref{PAFig}(b)). Consequently, it is not obvious that the onset of an LSP triggers mass loss.

An interesting insight is that, between $\log{P_{B+C^\prime}[W_{\rm JK}=12]} \approx 1.5$ and 1.6 dex, the projected periods of the first-overtone and LSP modes ($\log{P_{B+C^\prime}[W_{\rm JK}=12]}, \log{P_{D}[W_{\rm JK}=12]}$) anti-correlate for most stars (Figure \ref{DiscFig}(a); more precisely, they increase at different rates with luminosity). This suggests that stars evolve from the long- to short-period side of sequence $D$. Hence, stars on the short-period side of sequence $D$ are dust-producing, and stars on the long-period side are not (Figure \ref{PAFig}(c)), reflecting the transition from $B$ to $C^\prime$. A second group of stars forms a vertical sequence in Figure \ref{DiscFig}(a). Their origin is unknown, but they may represent spurious LSP periods.

The anti-correlation in Figure \ref{DiscFig}(a) could also be produced if the slopes of sequence $D$ and $B$+$C^\prime$ in the period--luminosity diagram differ slightly, i.e., if the period ratio of sequences $B$+$C^\prime$ and $D$ differ with stellar mass and/or gravity. However, this would not produce the decline in dust-producing stars with increasing sequence $D$ period, as seen in Figure \ref{PAFig}(c). Further analysis of these phenomena may help us understand the origin of the LSP modes.

As the LSP forms and stars transition between sequences $B \rightarrow C^\prime$, the fundamental mode (sequences $C$ and possibly $F$) starts to become prominent (Figure \ref{PAFig}(b)), and the amplitudes of the first overtone and fundamental modes and LSP all increase in a closely correlated fashion. This led us to investigate whether the sum of amplitudes correlated better with mass-loss rate than the amplitude of the primary mode. Indeed, a close correlation is found, with mass loss becoming significant when the combination of first-overtone and fundamental modes exceeds $\Delta I \approx 0.09$ mag (Figure \ref{DiscFig}(b)). Therefore, we suggest it is the strong amplification of one or both of these modes, possibly assisted by the LSPs, that trigger strong mass loss by levitating the star's outer layers.

Figure \ref{DiscFig}(c) shows the relationship between amplitude and mass-loss rate. Two populations can clearly be identified: one with $\Delta I_1 + \Delta I_2 \lesssim 0.09$ mag and essentially zero mass loss; and one with $\Delta I_1 + \Delta I_2 \gtrsim 0.09$ mag, with $\dot{M} \sim 10^{-7}$ M$_\odot$ yr$^{-1}$. A sparsely populated bridge connects these two populations. Hence, we appear to have a rapid transition between a quiescient state and a mass-losing state. That mass-losing state has largely fixed properties, suggesting some form of saturation in the energy transport: there is less than a factor of two increase in average mass-loss rate between $\Delta I = 0.09$ and 0.30 mag on sequence $C^\prime$, and $\Delta I = 0.09$ and 0.50 mag on sequence $C$. The average mass-loss rate from \citet{RSSM12} over these two ranges is $\log(\dot{M}) = -7.39$ [M$_\odot$ yr$^{-1}$], with a standard deviation of 0.42 dex. This consistency in mass-loss rate reflects what is seen in Galactic stars, where $\log(\dot{M}) = -6.43$ [M$_\odot$ yr$^{-1}$] was found by \citet{MDBZL18}. The factor of ten difference between the LMC and Galaxy is likely attributable to the different methods for calculating mass-loss rate, rather than reflecting any real difference. Since mass-loss rates from \citet{MDBZL18} are gas mass-loss rates, provided a gas:dust ratio of 400:1 is appropriate, the mass-loss rates of \citet{RSSM12} may need to be scaled up by a factor of $\sim$10.

Once amplitudes exceed $\Delta I \sim 0.5$ mag, mass-loss rate increases dramatically, reaching $\log(\dot{M}) \sim -6$ [M$_\odot$ yr$^{-1}$] by $\Delta I = 1.5$ mag (this is only reached on sequence $C$ for relatively massive stars). The trigger for this second increase isn't obvious, but again seems linked to increasing pulsation amplitude, and approximately corresponds to the point where \citet{BHAE15} begin to produce dust-driven winds in model atmospheres.


\section{Conclusions}
\label{ConcSect}

We have clearly tied the onset of strong mass loss ($\dot{M} \gtrsim 10^{-8}$ M$_\odot$ yr$^{-1}$) from AGB stars to their transition between pulsation sequences $B$ and $C^\prime$, both first-overtone pulsation modes. This co-incides with the onset of LSPs, as previously noted by \citet{WN09}. However, it appears to have a stronger link to the presence of significant power in the fundamental-mode pulsation (sequence $C$). It is the (possibly non-linear) combination of power in these pulsation modes (first-overtone, fundamental mode and possibly LSP) that appears to dictate when and whether a star transitions from a wind of $\dot{M} \sim 10^{-9}$ to $10^{-8}$ M$_\odot$ yr$^{-1}$ to $\dot{M} \sim 10^{-7}$ to $10^{-6}$ M$_\odot$ yr$^{-1}$. Most stars likely end their AGB evolution during this phase, with a relatively constant mass-loss rate. Only the most luminous, more massive stars undergo the AGB superwind, when $\dot{M} > 10^{-6}$ M$_\odot$ yr$^{-1}$.

We have also identified that, within each sequence, the LSP period anti-correlates with the fundamental or first-overtone period in the same star, and dust-production rate appears to increase from long to short periods. We interpret this as stars migrating from long-period LSPs to short-period LSPs over time, counter to other pulsation sequences.

To determine a new AGB mass-loss law, based on this information, we need a better calibration of how pulsation periods, amplitudes and modes relate to properties such as stellar temperature, luminosity and mass. We also need to investigate stars at differing metallicities, to determine how a star's composition affects the onset of its strong mass loss, and its mass-loss rate.


\section*{Acknowledgements}
While this article was written pro bono, IM acknowledges previous support for this research from the UK Science and Technology Facilities Council under grant ST/L00768/1. MT acknowledges support from the ERC Consolidator Grant funding scheme (\textit{project STARKEY}, G.A.\ n.\ 615604). We thank Albert Zijlstra and the anonymous referee for helpful comments on this paper.



\begin{thebibliography}{}
\makeatletter
\relax
\def\mn@urlcharsother{\let\do\@makeother \do\$\do\&\do\#\do\^\do\_\do\%\do\~}
\def\mn@doi{\begingroup\mn@urlcharsother \@ifnextchar [ {\mn@doi@}
  {\mn@doi@[]}}
\def\mn@doi@[#1]#2{\def\@tempa{#1}\ifx\@tempa\@empty \href
  {http://dx.doi.org/#2} {doi:#2}\else \href {http://dx.doi.org/#2} {#1}\fi
  \endgroup}
\def\mn@eprint#1#2{\mn@eprint@#1:#2::\@nil}
\def\mn@eprint@arXiv#1{\href {http://arxiv.org/abs/#1} {{\tt arXiv:#1}}}
\def\mn@eprint@dblp#1{\href {http://dblp.uni-trier.de/rec/bibtex/#1.xml}
  {dblp:#1}}
\def\mn@eprint@#1:#2:#3:#4\@nil{\def\@tempa {#1}\def\@tempb {#2}\def\@tempc
  {#3}\ifx \@tempc \@empty \let \@tempc \@tempb \let \@tempb \@tempa \fi \ifx
  \@tempb \@empty \def\@tempb {arXiv}\fi \@ifundefined
  {mn@eprint@\@tempb}{\@tempb:\@tempc}{\expandafter \expandafter \csname
  mn@eprint@\@tempb\endcsname \expandafter{\@tempc}}}

\bibitem[\protect\citeauthoryear{{Alard} et~al.,}{{Alard}
  et~al.}{2001}]{ABC+01}
{Alard} C.,  et~al., 2001, \mn@doi [ApJ] {10.1086/320440}, \href
  {http://adsabs.harvard.edu/abs/2001ApJ...552..289A} {552, 289}

\bibitem[\protect\citeauthoryear{{Bladh}, {H{\"o}fner}, {Aringer}  \&
  {Eriksson}}{{Bladh} et~al.}{2015}]{BHAE15}
{Bladh} S.,  {H{\"o}fner} S.,  {Aringer} B.,   {Eriksson} K.,  2015, \mn@doi
  [A\&A] {10.1051/0004-6361/201424917}, \href
  {http://adsabs.harvard.edu/abs/2015A\%26A...575A.105B} {575, A105}

\bibitem[\protect\citeauthoryear{{Boyer} et~al.,}{{Boyer}
  et~al.}{2017}]{BMG+17}
{Boyer} M.~L.,  et~al., 2017, \mn@doi [ApJ] {10.3847/1538-4357/aa9892}, \href
  {http://adsabs.harvard.edu/abs/2017ApJ...851..152B} {851, 152}

\bibitem[\protect\citeauthoryear{{Cannon}}{{Cannon}}{1967}]{Cannon67}
{Cannon} R.~D.,  1967, The Observatory, \href
  {http://adsabs.harvard.edu/abs/1967Obs....87..231C} {87, 231}

\bibitem[\protect\citeauthoryear{{Cutri} et~al.,}{{Cutri}
  et~al.}{2013}]{CWC+13}
{Cutri} R.~M.,  et~al., 2013, Technical report, {Explanatory Supplement to the
  AllWISE Data Release Products}

\bibitem[\protect\citeauthoryear{{Derekas}, {Kiss}, {Bedding}, {Kjeldsen},
  {Lah}  \& {Szab{\'o}}}{{Derekas} et~al.}{2006}]{DKB+06}
{Derekas} A.,  {Kiss} L.~L.,  {Bedding} T.~R.,  {Kjeldsen} H.,  {Lah} P.,
  {Szab{\'o}} G.~M.,  2006, \mn@doi [ApJ] {10.1086/508686}, \href
  {http://adsabs.harvard.edu/cgi-bin/nph-bib_query?bibcode=2006ApJ...650L..55D&db_key=AST}
  {650, L55}

\bibitem[\protect\citeauthoryear{{Fraser}, {Hawley}, {Cook}  \&
  {Keller}}{{Fraser} et~al.}{2005}]{FHCK05}
{Fraser} O.~J.,  {Hawley} S.~L.,  {Cook} K.~H.,   {Keller} S.~C.,  2005,
  \mn@doi [AJ] {10.1086/426749}, \href
  {http://adsabs.harvard.edu/cgi-bin/nph-bib_query?bibcode=2005AJ....129..768F&db_key=AST}
  {129, 768}

\bibitem[\protect\citeauthoryear{{Glass} \& {van Leeuwen}}{{Glass} \& {van
  Leeuwen}}{2007}]{GvL07}
{Glass} I.~S.,  {van Leeuwen} F.,  2007, \mn@doi [MNRAS]
  {10.1111/j.1365-2966.2007.11903.x}, \href
  {http://adsabs.harvard.edu/abs/2007MNRAS.378.1543G} {378, 1543}

\bibitem[\protect\citeauthoryear{{Glass}, {Schultheis}, {Blommaert}, {Sahai},
  {Stute}  \& {Uttenthaler}}{{Glass} et~al.}{2009}]{GSB+09}
{Glass} I.~S.,  {Schultheis} M.,  {Blommaert} J.~A.~D.~L.,  {Sahai} R.,
  {Stute} M.,   {Uttenthaler} S.,  2009, \mn@doi [MNRAS]
  {10.1111/j.1745-3933.2009.00628.x}, \href
  {http://adsabs.harvard.edu/abs/2009MNRAS.395L..11G} {395, L11}

\bibitem[\protect\citeauthoryear{{Goldman} et~al.,}{{Goldman}
  et~al.}{2017}]{GvLZ+17}
{Goldman} S.~R.,  et~al., 2017, \mn@doi [MNRAS] {10.1093/mnras/stw2708}, \href
  {http://adsabs.harvard.edu/abs/2017MNRAS.465..403G} {465, 403}

\bibitem[\protect\citeauthoryear{{Groenewegen}}{{Groenewegen}}{2014}]{Groenewegen14}
{Groenewegen} M.~A.~T.,  2014, \mn@doi [A\&A] {10.1051/0004-6361/201322671},
  \href {http://adsabs.harvard.edu/abs/2014A\%26A...561L..11G} {561, L11}

\bibitem[\protect\citeauthoryear{{Groenewegen} et~al.,}{{Groenewegen}
  et~al.}{2016}]{GVM+16}
{Groenewegen} M.~A.~T.,  et~al., 2016, \mn@doi [A\&A]
  {10.1051/0004-6361/201629590}, \href
  {http://adsabs.harvard.edu/abs/2016A\%26A...596A..50G} {596, A50}

\bibitem[\protect\citeauthoryear{{Habing}}{{Habing}}{1996}]{Habing96}
{Habing} H.~J.,  1996, \mn@doi [A\&AR] {10.1007/PL00013287}, \href
  {http://adsabs.harvard.edu/abs/1996A\%26ARv...7...97H} {7, 97}

\bibitem[\protect\citeauthoryear{{Herwig}}{{Herwig}}{2005}]{Herwig05}
{Herwig} F.,  2005, \mn@doi [ARA\&A] {10.1146/annurev.astro.43.072103.150600},
  \href {http://adsabs.harvard.edu/abs/2005ARA\%26A..43..435H} {43, 435}

\bibitem[\protect\citeauthoryear{{H{\"o}fner} \& {Olofsson}}{{H{\"o}fner} \&
  {Olofsson}}{2018}]{HO18}
{H{\"o}fner} S.,  {Olofsson} H.,  2018, \mn@doi [A\&AR]
  {10.1007/s00159-017-0106-5}, \href
  {http://adsabs.harvard.edu/abs/2018A\%26ARv..26....1H} {26, 1}

\bibitem[\protect\citeauthoryear{{Ita} et~al.,}{{Ita} et~al.}{2004}]{ITM+04}
{Ita} Y.,  et~al., 2004, \mn@doi [MNRAS] {10.1111/j.1365-2966.2004.07257.x},
  \href
  {http://adsabs.harvard.edu/cgi-bin/nph-bib_query?bibcode=2004MNRAS.347..720I&db_key=AST}
  {347, 720}

\bibitem[\protect\citeauthoryear{{Jones} et~al.,}{{Jones}
  et~al.}{2017}]{JWK+17}
{Jones} O.~C.,  et~al., 2017, \mn@doi [\mnras] {10.1093/mnras/stx1101}, \href
  {http://adsabs.harvard.edu/abs/2017MNRAS.470.3250J} {470, 3250}

\bibitem[\protect\citeauthoryear{{Karakas} \& {Lattanzio}}{{Karakas} \&
  {Lattanzio}}{2014}]{KL14}
{Karakas} A.~I.,  {Lattanzio} J.~C.,  2014, \mn@doi [PASA]
  {10.1017/pasa.2014.21}, \href
  {http://adsabs.harvard.edu/abs/2014PASA...31...30K} {31, e030}

\bibitem[\protect\citeauthoryear{{Lattanzio} \& {Wood}}{{Lattanzio} \&
  {Wood}}{2003}]{LW03}
{Lattanzio} J.~C.,  {Wood} P.~R.,  2003, in {Habing} H.~J.,  {Olofsson} H.,
  eds, Asymptotic giant branch stars, by Harm J. Habing and Hans Olofsson.
  Astronomy and astrophysics library, New York, Berlin: Springer, 2003, p. 23.
  p.~23

\bibitem[\protect\citeauthoryear{{Madore}}{{Madore}}{1982}]{Madore82}
{Madore} B.~F.,  1982, \mn@doi [ApJ] {10.1086/159659}, \href
  {http://adsabs.harvard.edu/abs/1982ApJ...253..575M} {253, 575}

\bibitem[\protect\citeauthoryear{{Matsuura} et~al.,}{{Matsuura}
  et~al.}{2016}]{MSS+16}
{Matsuura} M.,  et~al., 2016, \mn@doi [MNRAS] {10.1093/mnras/stw1853}, \href
  {http://adsabs.harvard.edu/abs/2016MNRAS.462.2995M} {462, 2995}

\bibitem[\protect\citeauthoryear{{McDonald} \& {Zijlstra}}{{McDonald} \&
  {Zijlstra}}{2015}]{MZ15b}
{McDonald} I.,  {Zijlstra} A.~A.,  2015, \mn@doi [MNRAS]
  {10.1093/mnras/stv007}, \href
  {http://adsabs.harvard.edu/abs/2015MNRAS.448..502M} {448, 502}

\bibitem[\protect\citeauthoryear{{McDonald} \& {Zijlstra}}{{McDonald} \&
  {Zijlstra}}{2016}]{MZ16}
{McDonald} I.,  {Zijlstra} A.~A.,  2016, \mn@doi [ApJ]
  {10.3847/2041-8205/823/2/L38}, \href
  {http://cdsads.u-strasbg.fr/abs/2016ApJ...823L..38M} {823, L38}

\bibitem[\protect\citeauthoryear{{McDonald}, {Boyer}, {van Loon}  \&
  {Zijlstra}}{{McDonald} et~al.}{2011}]{MBvLZ11}
{McDonald} I.,  {Boyer} M.~L.,  {van Loon} J.~T.,   {Zijlstra} A.~A.,  2011,
  \mn@doi [ApJ] {10.1088/0004-637X/730/2/71}, \href
  {http://adsabs.harvard.edu/abs/2011ApJ...730...71M} {730, 71}

\bibitem[\protect\citeauthoryear{{McDonald}, {Zijlstra}  \& {Boyer}}{{McDonald}
  et~al.}{2012}]{MZB12}
{McDonald} I.,  {Zijlstra} A.~A.,   {Boyer} M.~L.,  2012, \mn@doi [MNRAS]
  {10.1111/j.1365-2966.2012.21873.x}, \href
  {http://adsabs.harvard.edu/abs/2012MNRAS.427..343M} {427, 343}

\bibitem[\protect\citeauthoryear{{McDonald}, {Zijlstra}, {Sloan}, {Lagadec},
  {Johnson}, {Uttenthaler}, {Jones}  \& {Smith}}{{McDonald}
  et~al.}{2016}]{MZS+16}
{McDonald} I.,  {Zijlstra} A.~A.,  {Sloan} G.~C.,  {Lagadec} E.,  {Johnson}
  C.~I.,  {Uttenthaler} S.,  {Jones} O.~C.,   {Smith} C.~L.,  2016, \mn@doi
  [MNRAS] {10.1093/mnras/stv2942}, \href
  {http://adsabs.harvard.edu/abs/2016MNRAS.456.4542M} {456, 4542}

\bibitem[\protect\citeauthoryear{{McDonald}, {Zijlstra}  \&
  {Watson}}{{McDonald} et~al.}{2017}]{MZW17}
{McDonald} I.,  {Zijlstra} A.~A.,   {Watson} R.~A.,  2017, \mn@doi [MNRAS]
  {10.1093/mnras/stx1433}, \href
  {http://adsabs.harvard.edu/abs/2017MNRAS.471..770M} {471, 770}

\bibitem[\protect\citeauthoryear{{McDonald}, {De Beck}, {Zijlstra}  \&
  {Lagadec}}{{McDonald} et~al.}{2018}]{MDBZL18}
{McDonald} I.,  {De Beck} E.,  {Zijlstra} A.~A.,   {Lagadec} E.,  2018, \mn@doi
  [MNRAS] {10.1093/mnras/sty2607}, \href
  {http://adsabs.harvard.edu/abs/2018MNRAS.481.4984M} {481, 4984}

\bibitem[\protect\citeauthoryear{{McDonald}, {Boyer}, {Groenewegen}, {Lagadec},
  {Richards}, {Sloan}  \& {Zijlstra}}{{McDonald} et~al.}{2019}]{MBG+19}
{McDonald} I.,  {Boyer} M.~L.,  {Groenewegen} M. A.~T.,  {Lagadec} E.,
  {Richards} A. M.~S.,  {Sloan} G.~C.,   {Zijlstra} A.~A.,  2019, arXiv
  e-prints, \href {https://ui.adsabs.harvard.edu/\#abs/2019arXiv190105416M} {p.
  arXiv:1901.05416}

\bibitem[\protect\citeauthoryear{{Meixner} et~al.,}{{Meixner}
  et~al.}{2006}]{MGI+06}
{Meixner} M.,  et~al., 2006, \mn@doi [AJ] {10.1086/508185}, \href
  {http://adsabs.harvard.edu/abs/2006AJ....132.2268M} {132, 2268}

\bibitem[\protect\citeauthoryear{{Nicholls}, {Wood}, {Cioni}  \&
  {Soszy{\'n}ski}}{{Nicholls} et~al.}{2009}]{NWCS09}
{Nicholls} C.~P.,  {Wood} P.~R.,  {Cioni} M.,   {Soszy{\'n}ski} I.,  2009,
  \mn@doi [MNRAS] {10.1111/j.1365-2966.2009.15401.x}, \href
  {http://adsabs.harvard.edu/abs/2009MNRAS.399.2063N} {399, 2063}

\bibitem[\protect\citeauthoryear{{Riebel}, {Srinivasan}, {Sargent}  \&
  {Meixner}}{{Riebel} et~al.}{2012}]{RSSM12}
{Riebel} D.,  {Srinivasan} S.,  {Sargent} B.,   {Meixner} M.,  2012, \mn@doi
  [ApJ] {10.1088/0004-637X/753/1/71}, \href
  {http://cdsads.u-strasbg.fr/abs/2012ApJ...753...71R} {753, 71}

\bibitem[\protect\citeauthoryear{{Soszy{\'n}ski} \& {Wood}}{{Soszy{\'n}ski} \&
  {Wood}}{2013}]{SW13}
{Soszy{\'n}ski} I.,  {Wood} P.~R.,  2013, \mn@doi [ApJ]
  {10.1088/0004-637X/763/2/103}, \href
  {http://adsabs.harvard.edu/abs/2013ApJ...763..103S} {763, 103}

\bibitem[\protect\citeauthoryear{{Soszy{\'n}ski}, {Udalski}, {Kubiak},
  {Szyma{\'n}ski}, {Pietrzy{\'n}ski}, {Zebrun}, {Szewczyk}  \&
  {Wyrzykowski}}{{Soszy{\'n}ski} et~al.}{2004}]{SUK+04}
{Soszy{\'n}ski} I.,  {Udalski} A.,  {Kubiak} M.,  {Szyma{\'n}ski} M.,
  {Pietrzy{\'n}ski} G.,  {Zebrun} K.,  {Szewczyk} O.,   {Wyrzykowski} {\L}.,
  2004, Acta Astronomica, \href
  {http://adsabs.harvard.edu/cgi-bin/nph-bib_query?bibcode=2004AcA....54..129S&db_key=AST}
  {54, 129}

\bibitem[\protect\citeauthoryear{{Soszynski} et~al.,}{{Soszynski}
  et~al.}{2005}]{SUK+05}
{Soszynski} I.,  et~al., 2005, Acta~Astron., \href
  {http://adsabs.harvard.edu/abs/2005AcA....55..331S} {55, 331}

\bibitem[\protect\citeauthoryear{{Soszy{\'n}ski} et~al.,}{{Soszy{\'n}ski}
  et~al.}{2009}]{SUS+09}
{Soszy{\'n}ski} I.,  et~al., 2009, Acta~Astron., \href
  {http://adsabs.harvard.edu/abs/2009AcA....59..239S} {59, 239}

\bibitem[\protect\citeauthoryear{{Srinivasan}, {Sargent}  \&
  {Meixner}}{{Srinivasan} et~al.}{2011}]{SSM11}
{Srinivasan} S.,  {Sargent} B.~A.,   {Meixner} M.,  2011, \mn@doi [A\&A]
  {10.1051/0004-6361/201117033}, \href
  {http://cdsads.u-strasbg.fr/abs/2011A\%26A...532A..54S} {532, A54}

\bibitem[\protect\citeauthoryear{{Trabucchi}, {Wood}, {Montalb{\'a}n},
  {Marigo}, {Pastorelli}  \& {Girardi}}{{Trabucchi} et~al.}{2017}]{TWM+17}
{Trabucchi} M.,  {Wood} P.~R.,  {Montalb{\'a}n} J.,  {Marigo} P.,  {Pastorelli}
  G.,   {Girardi} L.,  2017, \mn@doi [ApJ] {10.3847/1538-4357/aa8998}, \href
  {http://adsabs.harvard.edu/abs/2017ApJ...847..139T} {847, 139}

\bibitem[\protect\citeauthoryear{{Trabucchi}, {Wood}, {Montalb{\'a}n},
  {Marigo}, {Pastorelli}  \& {Girardi}}{{Trabucchi} et~al.}{2019}]{TWM+19}
{Trabucchi} M.,  {Wood} P.~R.,  {Montalb{\'a}n} J.,  {Marigo} P.,  {Pastorelli}
  G.,   {Girardi} L.,  2019, \mn@doi [MNRAS] {10.1093/mnras/sty2745}, \href
  {http://adsabs.harvard.edu/abs/2019MNRAS.482..929T} {482, 929}

\bibitem[\protect\citeauthoryear{{Vassiliadis} \& {Wood}}{{Vassiliadis} \&
  {Wood}}{1993}]{VW93}
{Vassiliadis} E.,  {Wood} P.~R.,  1993, \mn@doi [ApJ] {10.1086/173033}, \href
  {http://adsabs.harvard.edu/abs/1993ApJ...413..641V} {413, 641}

\bibitem[\protect\citeauthoryear{{Willson}}{{Willson}}{2000}]{Willson00}
{Willson} L.~A.,  2000, \mn@doi [ARA\&A] {10.1146/annurev.astro.38.1.573},
  \href {http://adsabs.harvard.edu/abs/2000ARA\%26A..38..573W} {38, 573}

\bibitem[\protect\citeauthoryear{{Wood}}{{Wood}}{2015}]{Wood15}
{Wood} P.~R.,  2015, \mn@doi [MNRAS] {10.1093/mnras/stv289}, \href
  {http://adsabs.harvard.edu/abs/2015MNRAS.448.3829W} {448, 3829}

\bibitem[\protect\citeauthoryear{{Wood} \& {Nicholls}}{{Wood} \&
  {Nicholls}}{2009}]{WN09}
{Wood} P.~R.,  {Nicholls} C.~P.,  2009, \mn@doi [ApJ]
  {10.1088/0004-637X/707/1/573}, \href
  {http://adsabs.harvard.edu/abs/2009ApJ...707..573W} {707, 573}

\bibitem[\protect\citeauthoryear{{van Loon}, {Boyer}  \& {McDonald}}{{van Loon}
  et~al.}{2008}]{vLBM08}
{van Loon} J.~T.,  {Boyer} M.~L.,   {McDonald} I.,  2008, \mn@doi [ApJ]
  {10.1086/589711}, \href {http://adsabs.harvard.edu/abs/2008ApJ...680L..49V}
  {680, L49}

\makeatother
\end{thebibliography}



\bsp	
\label{lastpage}
\end{document}